\documentclass[sigconf,10pt,nonacm]{acmart}
\usepackage{algorithmic}
\usepackage{graphicx}
\usepackage{subcaption}
\usepackage{array,multirow}

\setcopyright{none}

\makeatletter
\g@addto@macro\UrlSpecials{\do\!{\newline}}
\makeatother 

\begin{document}

\title{Sim2Field: End-to-End Development of AI RANs for 6G}

\author{Russell Ford, Pranav Madadi,\\Daoud Burghal}
\affiliation{%
		\institution{Samsung Research America}
		\city{Mountain View}
		\state{CA}
		\country{USA}}
\email{{russell.ford, p.madadi, d.burghal}@samsung.com}

\author{Mandar Kulkarni, Xiaochuan Ma, Guanbo Chen, Yan Xin}
\affiliation{%
			\institution{Samsung Research America}
			\city{Berkeley Heights}
			\state{NJ}
			\country{USA}}
\email{{m.kulkarni2, shawn.ma, guanbo.h, yan.xin}@}
\email{samsung.com}

\author{Hao Chen, Yeqing Hu, Chance Tarver, Panagiotis Skrimponis, Vitali Loseu,\\Yu Zhang, Yang Li, Jianzhong Zhang}
\affiliation{%
			\institution{Samsung Research America}
			\city{Plano}
			\state{TX}
			\country{USA}}
\email{{hao.chen1, yeqing.h, c.tarver, notis.sk, v.loseu,!y3.zhang, yang.li1, jianzhong.z}@samsung.com}

\author{Shubham Khunteta, Yeswanth G. Reddy, Ashok K. R. Chavva,\\Mahantesh Kothiwale}
\affiliation{%
		\institution{Samsung Research India}
		\city{Bangalore}
		\country{India}}
\email{{sk.khunteta, guddeti.r, ashok.chavva, mahantesh.k}@}
\email{samsung.com}

\author{Davide Villa}
\affiliation{%
	\institution{Northeastern University}
	\city{Boston}
	\state{MA}
	\country{USA}}
\email{villa.d@northeastern.edu}


\renewcommand{\shortauthors}{R. Ford et al.}

\begin{abstract}
Following state-of-the-art research results, which showed the potential for significant performance gains by applying AI/ML techniques in the cellular Radio Access Network (RAN), the wireless industry is now broadly pushing for the adoption of AI in 5G and future 6G technology.  Despite this enthusiasm, AI-based wireless systems still remain largely untested in the field. Common simulation methods for generating datasets for AI model training suffer from “reality gap” and, as a result, the performance of these simulation-trained models may not carry over to practical cellular systems. Additionally, the cost and complexity of developing high-performance proof-of-concept implementations present major hurdles for evaluating AI wireless systems in the field. In this work, we introduce a methodology which aims to address the challenges of bringing AI to real networks. We discuss how detailed Digital Twin simulations may be employed for training site-specific AI Physical (PHY) layer functions. We further present a powerful testbed for AI-RAN research and demonstrate how it enables rapid prototyping, field testing and data collection. Finally, we evaluate an AI channel estimation algorithm over-the-air with a commercial UE, demonstrating that real-world throughput gains of up to 40\% are achievable by incorporating AI in the physical layer. 
\end{abstract}

%
%
%


\maketitle

\section{Introduction}
In recent years, the application of Artificial Intelligence (AI) and machine learning to the wireless networking domain has seen a surge of interest in the telecom industry, thanks to the potential for large performance gains. Beyond enhancing current 5G, AI is expected to revolutionize future 6G technology, particularly at the wireless Physical (PHY) layer, where proposed AI-based Channel Estimation (CE), detection, decoding and Channel State Information (CSI) compression methods, among others, have shown that impressive gains in spectral efficiency and throughput may be possible \cite{zhang_survey}. While these results are indeed promising, they generally rely on simulation for generating datasets for AI training and evaluation. Thus far, there have been few prototype implementations or field trials demonstrating the true effectiveness of AI at the PHY layer under real-world conditions. Therefore, it may be argued that many of the challenges of making these advanced algorithms perform well in the field have yet to be thoroughly investigated on a practical level. 
\begin{figure}[t]
	\includegraphics[width=0.7\columnwidth]{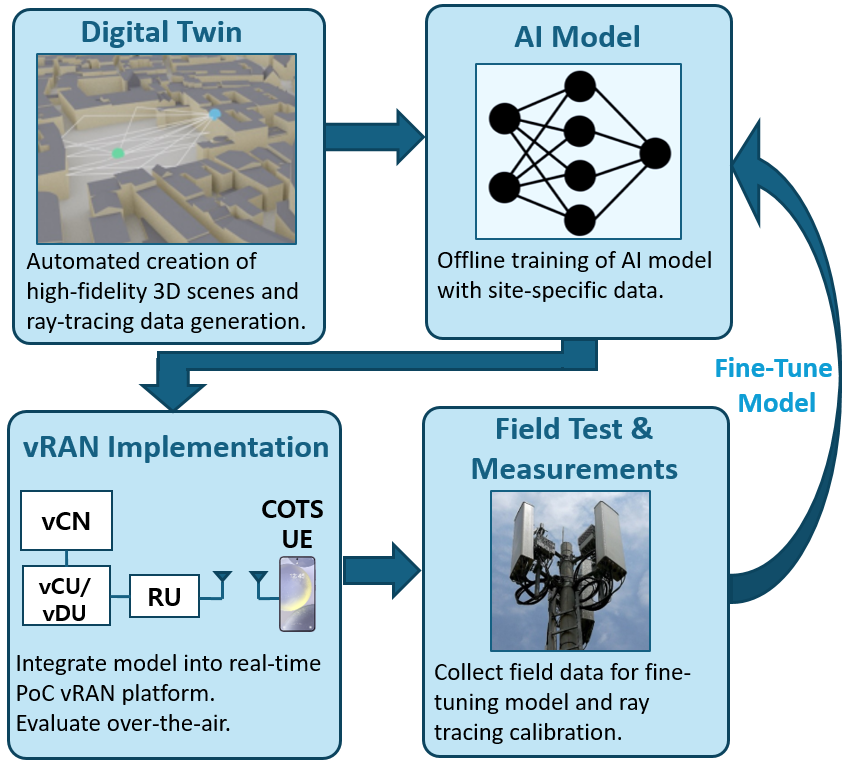}
	\setlength{\belowcaptionskip}{-14pt}
	\caption{Platform for rapid AI wireless innovation.}
	\label{fig:concept_diag}
\end{figure}

To accelerate the development of novel AI algorithms and ensure their reliability in real operator networks, a new development approach and experimental platform is required. In this work, we propose such a system, which we hope will serve as a framework for other researchers to rapidly innovate new 5G/6G technologies and address the many challenges of making AI RAN a commercial reality. The proposed framework is illustrated in Fig~\ref{fig:concept_diag}, which also describes the high-level structure of the paper.  

To summarize the key contributions of this work, 
\begin{enumerate}
	\item{In Section~\ref{sec:sim2field}, we introduce a ``Simulation-to-Field" (abbreviated Sim2Field) methodology, which aims to address the practical challenges of AI RAN development through site-specific, Digital Twin-based training. Insights are provided on how channel data may be synthesized to address the reality gap of standard channel models and improve data diversity.}
	\item{Methods are proposed for adding RF impairments to the generated channels, which is found to improve estimation accuracy by over 1 dB.}
	\item{In Section~\ref{sec:ai_chest}, we explain how the Sim2Field approach is applied toward the development of an AI uplink channel estimation algorithm. While the exemplary ResNet-based algorithm is not, by itself, a novel contribution, practical system integration aspects are presented in detail, which prior literature often neglects to address.}
	\item{In Section~\ref{sec:aerial_testbed}, we present an end-to-end prototype system based on the NVIDIA ARC-OTA platform. The ARC testbed enabled a real-time software implementation of the algorithm to be quickly brought up and tested Over-the-Air (OTA) using Commercial Off-the-Shelf (COTS) hardware.}
	\item{Lastly, in Section~\ref{sec:experiments}, we demonstrate the algorithm performance experimentally using a real-time channel emulator, as well as OTA in an indoor lab environment. Our AI-CE implementation achieves over a 40\% throughput gain in both OTA tests and channel emulator tests with mobility.}
\end{enumerate}

\subsection{Related Work}
A large body of work presently exists on machine learning and, in particular, deep learning methods for OFDM channel estimation \cite{soltani_chest, li_resnet_chest,cebed,gizzini_rnn_ce}. However, these studies are based on simulation results, and rarely discuss the practical hurdles at hand -- for instance, how to obtain accurate label data for training or compensate for RF impairments. In \cite{pena_mmwave_ce}, impairments such as phase noise and frequency offset are considered. However, purely theoretical models are employed, as opposed to the field data-driven approach of our work.
In \cite{deepsig_openran}, the authors propose a method for field data collection and fine tuning of a neural receiver model. Although tested with field data, OTA evaluation of the system with commercial devices is not conducted.
Several 5G testbeds have been developed by the wireless research community, notably POWDER \cite{powder}, COSMOS \cite{cosmos}, Colosseum \cite{colosseum}, O-RAN Gym \cite{oran_gym}, Open AI Cellular \cite{oaic}, and X5G \cite{X5G}. These systems are either limited in computing power, are not compliant with the 3GPP NR standard, or are designed mainly for evaluating higher-layer components, for example, O-RAN xApps, in contrast to baseband PHY components. Several works have developed prototypes and conducted OTA experiments to validate AI PHY algorithms but are simplistic and lack the real-time, NR-compliant capabilities to test with COTS UEs \cite{jiang_ai_ofdm, gizzini_rnn_ce}.

\section{Sim2Field Methodology}\label{sec:sim2field}
In wireless research today, AI models are commonly trained on datasets generated by stochastic channel models, which are popular thanks to their ease of implementation and simulation speed. These models are designed to characterize a general class of propagation environments, such as the Urban Microcell (UMi) and Rural Macrocell (RMa) scenarios defined by 3GPP \cite{3gpp_38901}. As such, they are not regarded as faithful representations of any specific real-world environment. The resulting “reality gap” thus poses a critical problem for developing AI models, as models pre-trained on simulated data may not generalize well and lose performance when deployed in the field.

While field measurements would provide the greatest accuracy in capturing the channel distribution, the cost and effort involved in collecting extensive measurements makes it infeasible for operators to conduct at scale. Ray Tracing (RT) simulation, on the other hand, presents a happy medium, as it represents the channel with a high degree of realism and without the need for costly measurements. Though known to be computationally-intensive, tools such as NVIDIA Sionna \cite{sionna} and Aerial Omniverse Digital Twin \cite{aodt} offer GPU acceleration of RT computation, allowing large datasets to be generated quickly.

Still, the Channel Impulse Responses (CIRs) generated by RT only reflect the propagation paths through the environment, as well as the Transmitter/Receiver (TX/RX) antenna patterns and Doppler effect, which may be explicitly modeled. However, the received I/Q samples also include impairments from the TX and RX components themselves. As a main contribution of this work, we explain how we have modeled these effects in the channel generation procedure to ensure the training data distribution is similar to the expected real distribution at the input of the baseband system.	

The proposed procedure for high-fidelity RT channel simulation of real-world environments, combined with receiver impairment modeling, lends to the development of a Digital Twin of the cellular network, as now described.

\subsection{Ray Tracing Scene Generation}\label{sec:ray_tracing}
Cellular Digital Twins are \textit{surrogate} models, which replicate the exact buildings, terrain, foliage and other blockages, as well as antennas and RF components, for cell sites in the operator network. To this end, we first generate 3D scenes for ray tracing from detailed map data, using tools such as Sionna and Blender \cite{blender} to automate most of the scene generation process.

The objects imported with the scene are then assigned materials based on their expected composition, such as concrete, glass and metal for buildings and earth for the ground plane. Each material type has associated electromagnetic (EM) properties, namely, permittivity, conductivity, and scattering coefficients, which are defined by the International Telecommunications Union (ITU) for various such materials. In our work, material assignment is done manually. However, though outside the scope of this paper, this could also be automated by AI image recognition techniques. 

The scene is then imported into the Sionna RT simulator and the NR gNBs and User Equipment (UE) devices are defined, along with their respective TX power and antenna array parameters, i.e. number of elements, polarization and radiation pattern. The gNB Radio Unit (RU) locations and respective parameters may be obtained from site engineering data for the operator network. UEs are randomly dropped in pre-defined regions within the scene. Ray shooting and bouncing is performed to compute paths between the TX and RX, taking into account scattering and diffraction at the given carrier frequency. Doppler shift is applied on the paths, which is a function of the UE velocities. The Channel Impulse Response (CIR) is then computed from these paths, which may be transformed to a Channel Frequency Response (CFR) $H(f)$ for the Orthogonal Frequency-Division Multiplexing (OFDM) channel at the subcarrier spacing defined for the specified NR numerology.

\begin{figure}[t]
	\includegraphics[width=0.8\columnwidth]{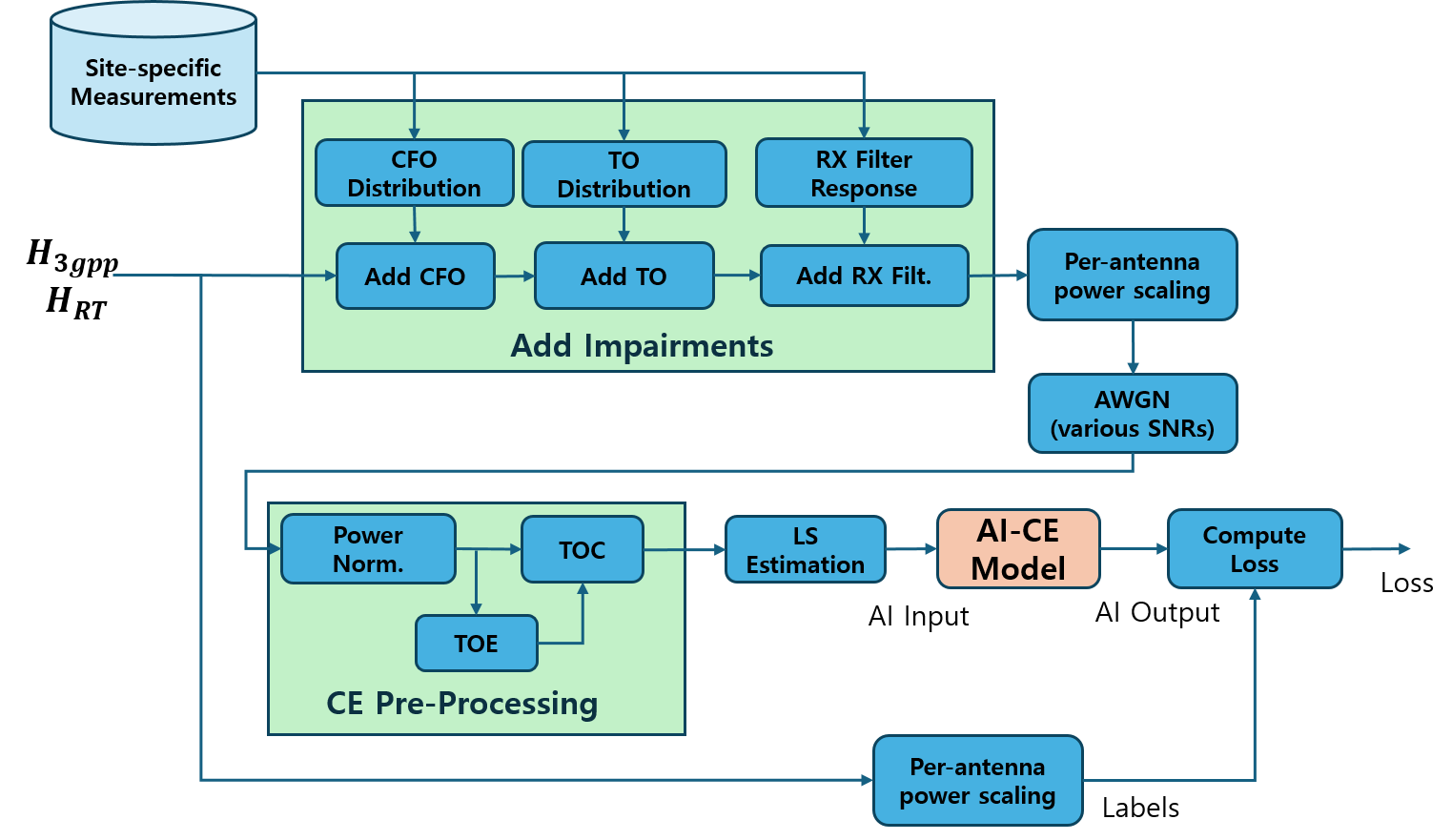}
	\setlength{\belowcaptionskip}{-10pt}
	\caption{Sim2Field training data generation procedure.}
	\label{fig:sim2field_training}
\end{figure}

\subsection{Domain Randomization and RT Calibration}
Regardless of how detailed one constructs a Digital Twin, inevitably some residual reality gap is expected with respect to the real environment. 
Techniques such as domain randomization, which creates diversity in the training data distribution by introducing random variations to the original dataset, have helped to overcome this dilemma in the robotics field and may prove useful for training AI wireless algorithms, as well. 


One approach is to randomize the locations of mobile devices and introduce scene objects of varying dimension and location, for example, by introducing random blockages in order to represent variation in foliage, vehicles, etc. The aforementioned material properties of different surfaces may also be varied. Although not a focus of this work, received power measurements at various locations may also be ascertained from Minimization of Drive Test (MDT) data. This would permit calibration of path loss and material properties, for example, by using techniques such as differentiable RT, to further improve simulation fidelity \cite{diffrt}.

\subsection{Receiver Impairment Modeling}
The RF front-end imparts numerous non-idealities to the received signal, including Timing Offset (TO), Carrier Frequency Offset (CFO), Power Amplifier (PA) non-linearity, Phase Noise and Local Oscillator (LO) leakage. Additionally, the receiver will not have a flat frequency response, especially over the wide 100 MHz carrier bandwidth of 5G systems. 
Automatic Gain Control (AGC) also dynamically scales the signal in response to the time-varying received power. It is thus necessary to capture these effects in the channel generation procedure to improve the robustness of the trained AI model. How the key impairments are represented and applied for training data generation is summarized in Fig.~\ref{fig:sim2field_training} and described in the following.

\subsubsection*{Timing and Carrier Frequency Offset}
TO and CFO are often present in the received channel, the former due to imperfect synchronization, with the latter attributed to LO mismatch. While these impairments are explicitly compensated for in the baseband processing, residual time and frequency offset may still remain. To model these effects, measurements are obtained using the NR testbed system, described later in Section~\ref{sec:aerial_testbed}, for the uplink channel between the COTS RU and UE. Channel estimates are computed from the Demodulation Reference Symbols (DMRS) of the Physical Uplink Shared Channel (PUSCH) using a variant of the well-known Minimum Mean-Squared Error (MMSE) algorithm, described in \cite{mmse_pdp}. 
TO manifests in the frequency domain as a phase shift, which may be estimated directly from the CFRs by finding the average phase difference between adjacent pilot subcarriers $k, k+1$, expressed as 
\begin{equation}
TO^i = \frac{1}{N_{ant}(N_{p}-1)}\sum_{j=0}^{N_{ant}-1}\sum_{k=0}^{N_{p}-2} H_{est}^{i,j}(k) H_{est}^{i,j}(k+1)^{*}
\end{equation}
for symbol index $i$ within the set of measured CFRs, where $N_{ant}$ and $N_{p}$ denote the number of antenna ports and pilot symbols, respectively. Here, averaging over antenna ports $j$ improves estimation accuracy. Similarly, CFO may be estimated as
\begin{equation}
CFO^i = \frac{1}{N_{ant}N_{p}}\sum_{j=0}^{N_{ant}-1}\sum_{k=0}^{N_{p}-1} H_{est}^{i_1,j}(k) H_{est}^{i_2,j}(k)^{*}
\end{equation}
for DMRS symbols $i_1, i_2$ within the same TDD slot. As shown in Fig.~\ref{fig:sim2field_training}, the distribution of TO and CFO, estimated from measurement data, is then randomly sampled and applied to the set of RT channels $H_{RT}$, or may similarly be applied to channel responses generated from stochastic models, denoted $H_{3GPP}$.

\subsubsection*{RX Filter Response Scaling and LO Leakage}
Typically, the hardware vendor will have characterized the complex filter response of the RU, making it straightforward to incorporate into the channel generation procedure. Otherwise, it may be characterized by averaging the measured frequency-domain channel gain $H_{est}^{i,j}(k)$ over many symbols $i$. In this work, this was accomplished by collecting measurements in an anechoic chamber for a fixed RU and UE position, since the wireless channel is mostly stationary. A magnitude response with passband ripple of about $\pm 0.5$ dB was observed for the Foxconn 7800E RU used in our testbed. 
The set of RT channels $H_{RT}$ is then scaled by the estimated filter response.

Additionally, LO leakage is commonly found at the center frequency, referred to as the ``DC'' subcarrier. In our experimental platform, a strong DC signal was observed along with other unwanted mixer products at specific subcarrier indices.
When pre-processing pilot data for AI-CE inference, described further in Section~\ref{sec:aerial_testbed}, we replace the complex channel gain at the distorted subcarriers $k_{DC}$ with the average of its adjacent subcarrier gains at $k_{DC}-1, k_{DC}+1$, under the assumption that the channel is mostly flat over 3 consecutive pilots.

\subsubsection*{Per-Antenna Power Scaling}
For our RU device, it was found experimentally that the distributions of I/Q sample power per each antenna port varied widely, with the median power differing by as much as 1.5 dB between antennas. Also, this relative difference was observed to be almost constant over received OFDM symbols collected during the test. We therefore introduce a per-antenna power scaling factor in the channel dataset to mirror the behavior of the RU.

\subsubsection*{Additive Noise}
The Signal-to-Interference-plus-Noise Ratio (SINR) of the received signal can naturally vary with different path loss and inter-cell interference, on top of the receiver noise floor. The final step is to add Gaussian noise for a range of SINRs, over which the receiver operation is expected. 



\section{AI Channel Estimation} \label{sec:ai_chest}
Channel estimation is a key PHY-layer function for which data-driven, AI techniques may provide an improvement over their conventional counterparts. 
Previous work has shown that deep learning methods such as denoising CNNs, borrowed from the image processing domain, may offer gains compared to methods like MMSE estimation, which is widely used in 5G systems today \cite{saikrishna_chest}. 

\begin{figure}[t]
	\includegraphics[width=0.85\columnwidth]{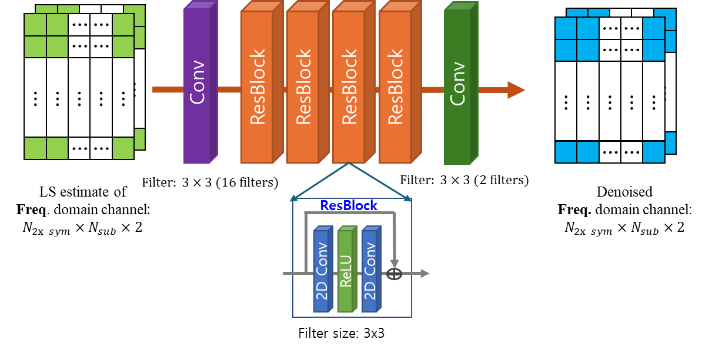}
	\setlength{\belowcaptionskip}{-10pt}
	\caption{Example ResNet CE model architecture.}
	\label{fig:resnet_model}
\end{figure}

Although the main contribution of this paper is not to introduce a novel AI-CE algorithm, in this section we present an exemplary AI model for uplink (PUSCH) CE, which is designed to be integrated in the real-time, NR-compliant platform. We base the model on the now established ResNet architecture thanks to its robust performance for image de-noising problems \cite{li_resnet_chest}. Illustrated in Fig.~\ref{fig:resnet_model}, the model input is a 3D tensor with dimensions $2N_{sym} \times N_{p} \times N_{ant}$, where $N_{sym}$ is the maximum number of OFDM symbols containing DMRS, $N_{p}$ is the maximum number of pilot subcarriers (1638 for a 100 MHz system) and $N_{ant}$ is the number of RX antennas of the RU. The first dimension corresponds to the CNN channels, for which there are two channels per DMRS symbol, since the complex channel gains are split into real and imaginary components. The input tensor contains the frequency-domain Least Squares (LS) estimates, which must be computed prior to AI estimation, as shown in Fig.~\ref{fig:sim2field_training}. The output is then the de-noised version of the LS estimates. An architecture with 4 ResBlocks is chosen to provide a balance between complexity and performance. 

\subsubsection*{Dynamic Resource Allocation}
One consideration for real systems, which is not typically addressed in the current AI wireless literature, is that many data structure dimensions must be able to scale dynamically, e.g., as different resources are assigned to each user. In the case of channel estimation, the number of DMRS symbols, Resource Blocks (RBs) and utilized antenna ports are all determined on a per-slot basis by the MAC scheduler. On the other hand, neural network input/output dimensions are generally fixed. 

In the PoC implementation introduced in Section~\ref{sec:aerial_testbed}, the dimensions $N_{ant}$ and $N_{sym}$ are fixed at 2 antennas and 3 symbols, respectively, due to limitations of the RU and MAC layer code. However, the set of occupied pilot subcarriers may vary according to the number of the scheduled RBs in each slot. To ensure that the AI-CE model is robust to input data tensors with differing dimensions, one may consider training multiple models with different configurations and dynamically switch between them at inference time. However, this would lead to increased system complexity and resource utilization of the target device. In this work, our approach to handling dynamic resource allocation, resulting in varying occupied data indices within the input tensor, was to train a single model by randomly zeroing-out ranges of indices during training. The loss function is then computed based on the output data tensor indices corresponding to the non-zero input indices only.

\begin{table}[b]
	\caption{Impact of RF impairments on AI-CE accuracy.}
	\begin{center}
		\begin{tabular}{|c|c|c|c|c|}
			\hline
			\multirow{2}{*}{\textbf{SNR (dB)}} & \multicolumn{4}{c|}{\textbf{NMSE (dB) for Test Case}} \\
			\cline{2-5}
			& \textbf{All} & \textbf{TO Off} & \textbf{Filt. Off} & \textbf{Scaling Off} \\
			\hline
			-10 & -10.35 & -9.56 & -10.31 & -9.70 \\
			\hline
			-5 & -15.49 & -14.98 & -15.32 & -15.31 \\
			\hline
			0 & -20.18 & -19.95 & -20.00 & -20.06 \\
			\hline
			5 & -24.40 & -24.15 & -24.18 & -24.25 \\
			\hline
			10 & -27.50 & -27.10 & -27.27 & -27.23  \\
			\hline
			15 & -29.25 & -28.67 & -28.99 & -28.80 \\
			\hline
			20 & -30.05 & -29.32 & -29.72 & -29.43 \\
			\hline
			Max. Impact & N/A & 0.79 & 0.33 & 0.65 \\
			\hline	
		\end{tabular}
		\label{tab:ablation_study}
	\end{center}
	\vspace{-2mm}
\end{table}

\subsection*{RF Impairment Ablation Study}
An ablation study was conducted to assess the impact of each aforementioned RF effect on AI-CE, the results of which are shown in Table~\ref{tab:ablation_study}. For each test case, a separate AI-CE model was trained with 8500 samples of simulated 3GPP CDL channel data. In the case denoted "All" in the table, all of the impairment models from the previous section are included. For the "TO Off," "Filt. Off," and "Scaling Off" cases, the timing offset modeling, RX filter gain modeling, and per-antenna power scaling were respectively disabled in the training data generation but were included in the test set. The channel estimation accuracy in terms of Normalized Mean-Squared Error (NMSE) w.r.t. the ground truth channel was then recorded for each case. The maximum impact, in terms of dB difference, of selectively disabling each impairment model is noted in the table. It was found that TO modeling and per-antenna scaling contributed the most to improve accuracy, with the RX filter model having less of a benefit.

\section{Real-Time vRAN Testbed}\label{sec:aerial_testbed}
Compared to traditional wireless signal processing algorithms and systems, for which the practical performance is well-established, the early stage of AI wireless technology makes it imperative for prototypes to be built for field evaluation. Unfortunately, due to the demands of 5G and future 6G RANs for high throughput and low latency, and the complexity of today's cellular standards, it is challenging to develop prototype systems that can approximate the capabilities of commercial base stations. Until recently, this has usually driven the need for high-speed FPGA or ASIC implementations, which are costly and involve long development cycles. 

\begin{figure}[t]
	\includegraphics[width=0.8\columnwidth]{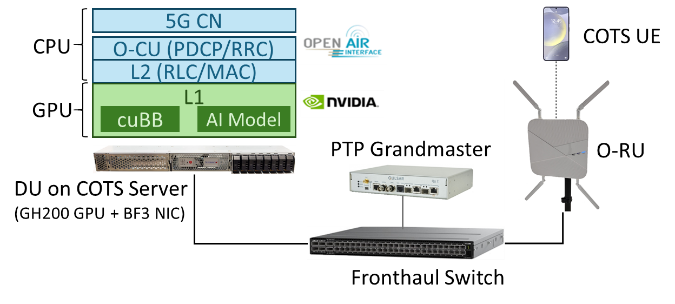}
	\setlength{\belowcaptionskip}{-10pt}
	\caption{Aerial ARC-OTA testbed.}
	\label{fig:aerial_testbed}
\end{figure}

We adopt the Aerial RAN CoLab Over-The-Air (ARC-OTA) system, which is a software-defined virtual RAN platform  developed by NVIDIA and the OpenAirInterface (OAI) Software Alliance (OSA). The testbed, shown in Fig.~\ref{fig:aerial_testbed}, combines powerful GPU servers for the base station Data Unit (DU) and commercial RUs for OTA transmission. OAI provides the Layer 2 MAC/RLC and Layer 3 functions, as well as the 5G core network elements. The L1 PHY functions leverage the NVIDIA Aerial CUDA-Accelerated RAN, referred to as cuBB, for efficient in-line GPU processing. Importantly, the CUDA/C++ software implementation of the L1 shortens the time for custom algorithms, like AI channel estimation, to be developed and tested over-the-air. Further details of the ARC-OTA platform are presented in \cite{kelkar_arc}.



\subsection*{Real-Time AI-CE Implementation in cuBB}
Another contribution of this work is the development of a framework for converting Python-based AI models to CUDA/C++ inference engines and, in turn, incorporating them into the L1 processing of the DU. 
The PUSCH computation in cuBB is implemented as a CUDA graph. This computational graph-based implementation requires minimal host synchronization, enabling highly-efficient use of both GPU and CPU resources. This makes it possible to run up to 20 cells with 100 MHz bandwidth and 4T/4R MIMO in real-time on a single DU server.

In this work, extensive modifications were made to the cuBB PUSCH implementation to integrate the AI model introduced in the earlier section. The NVIDIA TensorRT toolchain is used to convert the model, originally developed in Python/PyTorch, to an execution format that is compatible with CUDA/C++. Specifically, a TensorRT \textit{engine} is generated from the PyTorch model which, using the CUDA graph API, is then converted into a graph, with the underlying kernel functions generally corresponding to the layers of the CNN. This graph may then merged in as a sub-graph of the full cuBB PUSCH graph. The best possible latency and throughput performance is hence achieved for the CE module executing on the GPU.

\begin{figure}[t]
	\includegraphics[width=0.9\columnwidth]{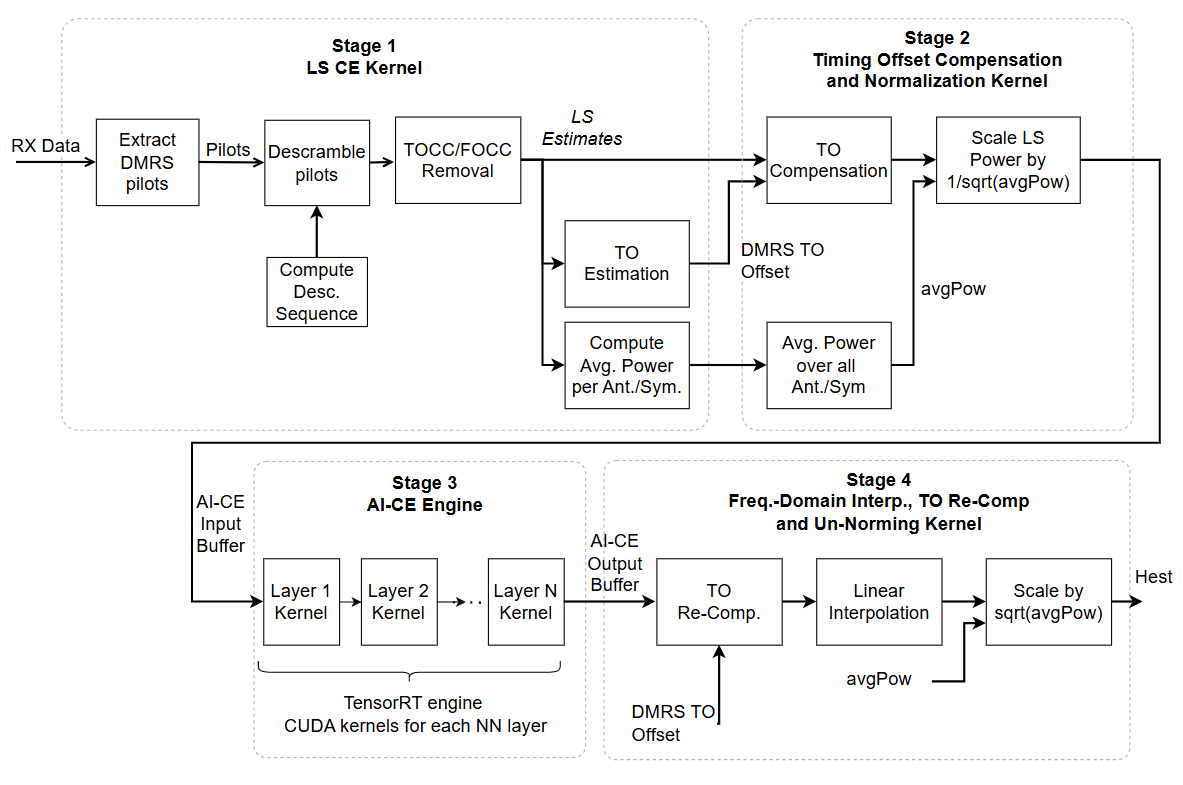}
	\setlength{\belowcaptionskip}{-10pt}
	\caption{AI-CE cuBB (CUDA) implementation.}
	\label{fig:cuda_ce_module}
\end{figure}

Fig.~\ref{fig:cuda_ce_module} shows the design of the AI-CE implementation in the cuBB code. For each OFDM symbol, the received frequency-domain IQ data samples are input to the “Stage 1” kernel, which performs DMRS sequence extraction, de-scrambling and Orthogonal Cover Code (OCC) removal. The result is a set of LS estimates, from which the TO may be estimated along with the average power. In the Stage 2 kernel, TO compensation and normalization of the AI model input signal is performed, which is then input to the AI-CE TensorRT engine, shown as Stage 3 in the diagram. The result of AI-CE inference is a tensor of channel estimates at pilot locations, which are passed to the Stage 4 kernel for TO re-compensation, de-normalization, and frequency-domain interpolation to compute the estimates at non-pilot subcarriers. The result is then passed to the channel equalizer module.

\section{Experimental Setup and Results}\label{sec:experiments}

In this section, we conduct experiments using the ARC-OTA system toward validating whether the large gains found in previous studies, for example \cite{li_resnet_chest}, are reasonable to expect in practical systems. In the sequel, we compare the performance of the proposed AI-CE algorithm, trained using the Sim2Field methodology, with respect to conventional MMSE. The baseline MMSE algorithm implemented in the Aerial cuBB PHY is based on the Power-Delay Profile (PDP) approximation scheme discussed in \cite{mmse_pdp}. Salient parameters pertaining to the emulator and OTA experiments are given in Table~\ref{tab:testbed_params}.

\subsection{Channel Emulator Experiments}
In this experiment, the performance of AI-CE is evaluated using a Keysight Propsim channel emulator. A custom AI-CE model is trained with 20k samples of statistical channel data from the well-known 3GPP Clustered Delay Line (CDL) channel model, specifically CDL-C and CDL-D from \cite{3gpp_38901}, with delay spread (DS) ranging from 10 to 100ns. The channel emulator is configured with the CDL-B channel with 300ns DS. The UE moves at a constant speed of 30 kmph along a circular path around a single RU with fixed position and antenna orientation, maintaining a constant distance of 200m from the RU. The UE velocity is thus kept constant w.r.t. the RU. In terms of physical setup, the RU is cabled directly to the input/output ports of the channel emulator. Additionally, a Samsung S24 smartphone is modified such that the device's antenna ports are hard-wired to cables with SMA connectors, allowing it to also be directly connected to the emulator. 

\begin{table}[t]
	\caption{Testbed Experimental Parameters}
	\begin{center}
		\begin{tabular}{|c|c|c|}
			\hline
			& \multicolumn{1}{c|}{\textbf{Parameter}} & \multicolumn{1}{c|}{\textbf{Value}} \\
			\hline
			\parbox[t]{2mm}{\multirow{10}{*}{\rotatebox[origin=c]{90}{General}}} & Carrier Frequency & 3.75 GHz (band n78) \\
			\cline{2-3}
			& Bandwidth & 100 MHz \\
			\cline{2-3}
			& Duplexing & TDD, DDDSU pattern \\
			\cline{2-3}
			& NR Numerology & $\mu=1$ (30 kHz SCS) \\
			\cline{2-3}
			& RU/UE Max. TX Power & 24 dBm \\
			\cline{2-3}
			& RU Antenna & 4 vertical dipoles \\
			\cline{2-3}
			& Num. DMRS symbols/slot & 3 \\
			\cline{2-3}
			& Max. PRB allocation & 273 \\
			\cline{2-3}
			& Max. UL Layers & 1 \\
			\cline{2-3}
			& Target BLER & 10\% \\
			\cline{2-3}
			\hline
			\parbox[t]{2mm}{\multirow{4}{*}{\rotatebox[origin=c]{90}{Emulation}}} & Channel Model & 38.901 CDL-B, 300 ns DS \\
			\cline{2-3}
			& UE Speed & 30 k/h \\
			\cline{2-3}
			& RU/UE Antenna Spacing & $0.5\lambda$ \\
			\cline{2-3}
			& Target SINR & -4 to 4 dB\\
			\hline
			\parbox[t]{2mm}{\multirow{2}{*}{\rotatebox[origin=c]{90}{OTA}}} & RU-UE Distance & 15 m\\
			\cline{2-3}
			& Target SINR & 5 dB\\
			\hline
		\end{tabular}
		\label{tab:testbed_params}
	\end{center}
	\vspace{-2mm}
\end{table}

\begin{figure}[t]
	\centering
	\begin{subfigure}{.5\columnwidth}
		\includegraphics[width=1.1\columnwidth]{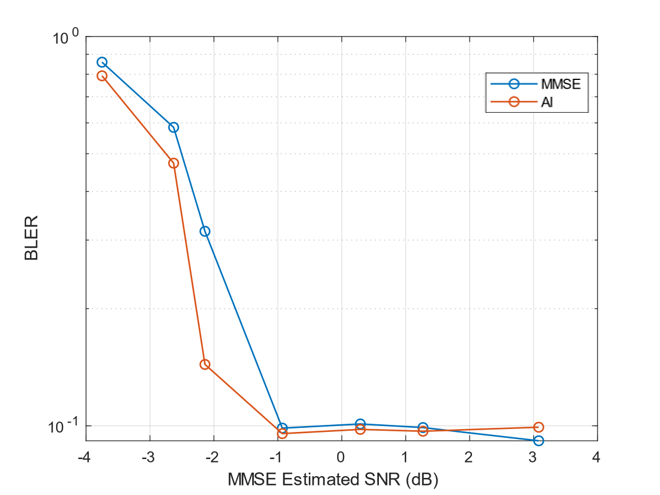}%
		\setlength{\belowcaptionskip}{-10pt}
		\caption{BLER \% vs. SNR}%
		\label{fig:emulator_bler}%
	\end{subfigure}\hfill%
	\begin{subfigure}{.5\columnwidth}
		\includegraphics[width=1.1\columnwidth]{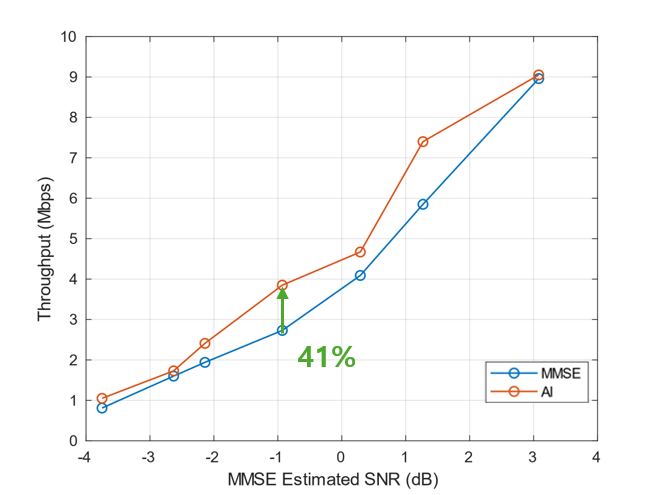}%
		\setlength{\belowcaptionskip}{-10pt}
		\caption{Throughput vs. SNR}%
		\label{fig:emulator_tput}%
	\end{subfigure}
	\setlength{\belowcaptionskip}{-5pt}
	\caption{Results for emulator experiments.}
	\label{fig:emulator_results}
\end{figure}

The testbed implementation enables end-to-end metrics, like UL Block Error Rate (BLER) and MAC-layer throughput, to be obtained. The mean of the observed throughput and BLER are plotted in Fig.~\ref{fig:emulator_results} over a range of SNRs. The SNR is swept by changing the UE TX power, which is controlled by setting the target SINR in the DU configuration. Closed-loop power control commands from the DU trigger the UE to adjust its TX power to meet the target SINR. Sixty seconds of data are collected for each SNR point.

As seen in the figure, appreciable gains in BLER and throughput of AI-CE w.r.t. MMSE are found over the range of SINRs between -2 to 2 dB, with up to a 41\% gain around -1 dB. Above 2-3 dB SINR, the performance of the two algorithms tends to converge. This suggests that AI-based channel estimation could be especially beneficial for cell edge users.
The result also gives evidence that an AI model trained with statistical channel data from one distribution (CDL-C/D, in this case) can still generalize and perform well on the unseen CDL-B channel distribution.

\begin{figure}[b]
	\includegraphics[width=0.6\columnwidth]{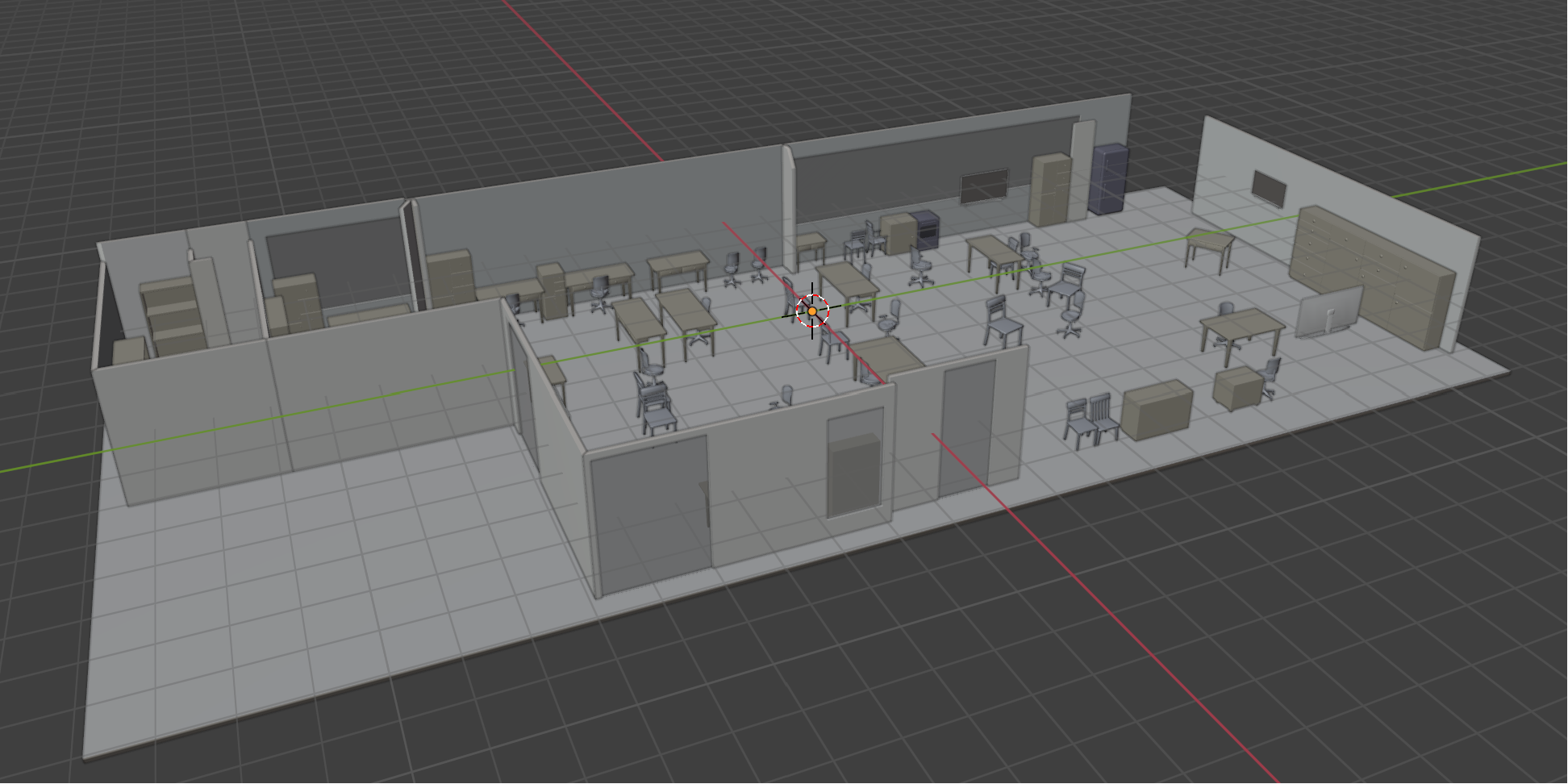}
	\caption{Blender model of lab for ray tracing.}
	\label{fig:lab_3dmodel}
\end{figure}

\subsection{Over-the-Air Experiments}
\begin{figure}[t]
	\includegraphics[width=0.9\columnwidth]{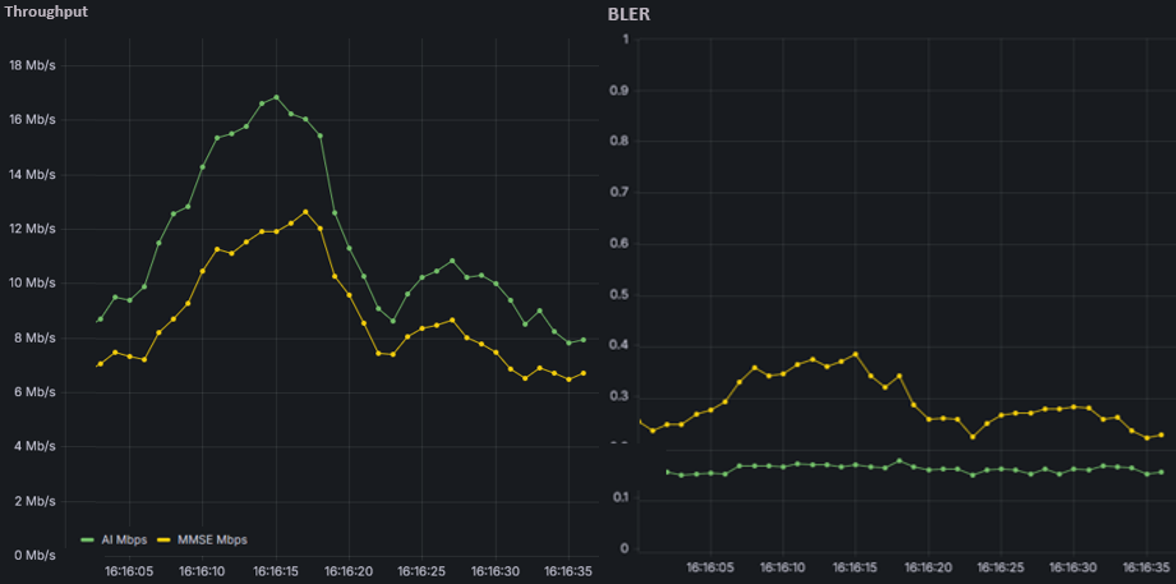}
	\setlength{\belowcaptionskip}{0pt}
	\caption{Dashboard with OTA tput. and BLER of AI vs. MMSE.}
	\label{fig:ota_grafana}
\end{figure}

Further experiments are conducted to assess the performance of AI-CE over-the-air in an indoor lab space. A 3D model of the lab, shown in Fig.~\ref{fig:lab_3dmodel} is created in Blender and imported into Sionna RT for channel data generation, again using the approach from Sec.~\ref{sec:sim2field} to train a site-specific model. During OTA testing, it was found that interference and movement of persons through the lab resulted in large variations in the channel, making it difficult to collect repeatable results to compare each algorithm. Modifications were made in the implementation to allow for multiple PUSCH processing chains to be instantiated for the same attached UE, in order to compare the real-time performance of AI-CE and the baseline in parallel with the same received I/Q data. The screenshot of the live dashboard in Fig.~\ref{fig:ota_grafana} shows that AI consistently outperforms MMSE by up to a 40\% margin in throughput. In this test, the UE is placed in a non-Line-of-Sight location 15m away from the RU and the target SINR is set to 5 dB.

In a separate experiment, an AI-CE model trained on 3GPP channel data is evaluated, along with a model trained on a combination of 3GPP and RT data from other scenes. Table~\ref{tab:ota_ai_rt_results} shows the comparison of these two models with the previous AI model trained on site-specific RT data, for the same OTA test setup. It is found that training on non-site specific RT data provides an additional gain over the 3GPP data-trained model, with a larger improvement coming from site-specific training. 

\begin{table}[t]
	\caption{Results for AI-CE trained with 3GPP vs. RT data.}
	\begin{center}
		\begin{tabular}{|c|c|c|}
			\hline
			\textbf{Model} & \textbf{BLER (\%)} & \textbf{Tput. (Mbps)} \\
			\hline
			AI-3GPP	& 18 & 19 \\
			\hline
			AI-RT (Non-site specific) & 16 & 20	\\
			\hline
			AI-RT (Site-specific) & 16 & 22 \\
			\hline
		\end{tabular}
	\label{tab:ota_ai_rt_results}
	\end{center}
	\vspace{-5mm}
\end{table}	

\section{Conclusions}
These early results encouragingly suggest that real-world gains are indeed possible for AI-based channel estimation and other PHY algorithms. However, to guarantee the robustness of AI algorithms in the field, more is required beyond naively training models from statistical simulation data. To our knowledge, this work is one of the first to comprehensively address many of the real-world aspects of AI PHY training. Still, before AI may be adopted in production systems, further innovation and testing is undoubtedly needed to ensure reliable performance for a broad range of network devices and environments.

\pagebreak

\bibliographystyle{ACM-Reference-Format}
\bibliography{references}

\end{document}